# Privacy Risks from Public Data Sources

*Zacharias Tzermias* FORTH-ICS, Greece
*Panagiotis Papadopoulos* FORTH-ICS, Greece
*Sotiris Ioannidis* FORTH-ICS, Greece
*Vassilis Prevelakis* IDA, TUBS, Germany

## 1. Introduction

Pavlos' story is very commonplace and highly predictable: his phone rung early one morning and the person who called identified herself as the representative of a well known national bank. She then asked Pavlos to confirm some personal information such as his tax registration number, address, identification card number, and so on. When Pavlos confirmed that all the provided information was correct, she told him that for identification purposes she needed his credit card number, expiration date and a three digit number printed on the back. He later told the police that since the person on the phone knew so much about him, he assumed that she was who she claimed to be. In fact she was a fraudster trying to extract information from Pavlos which was used to purchase airline tickets using Pavlos' credit card.

But how did the person on the phone know so many personal details about Pavlos? In most likelihood, she got it from the Greek government. This is so, because, the Greek government, like many other governments and private concerns, has been amassing personal data in electronic form since the beginning of the computer era. In the past two decades, the prevalence of the Internet and the technological advances in storage technologies have enabled an unprecedented volume of personal data to be collected.

While this data was stored in magnetic tapes sitting in silent basements, or even in large mainframes disconnected from the rest of the world, the risk for large scale loss of personal information was kept to a minimum. Each agency or company had its own bit of knowledge about Pavlos, but all these fragments remained disconnected. However, as all these information repositories become connected to the Internet and acquired front-ends that allow individuals on-line access to their data., they opened up the possibility that someone will be able to piece together all these fragments collected from different sources, constructing complete profiles of individuals like Pavlos, bypassing any privacy protection provided by law. For example, if two repositories store one's social security number (SSN) or credit card number but the first repository hides the first half while the other the last half, then by combining the two records, the entire number can be assembled. This is exactly what happened in the case of Mat Honan [1] where attackers combined his personal data maintained by Google, Amazon and Apple Store to reset his password and wipe all the data he kept in his accounts. The attackers even managed to remotely wipe clean his iPhone.

Even though such incidents are scary enough, it is our opinion that a greater risk lies in the collection of data by the state, or by organizations and agencies under state control. We believe that state administered data repositories constitute a greater risk for individuals, because (a) submitting data requested by the government is usually obligatory and, in any event, if the state wants access to personal data, they can get to it, even if they have to change the law; (b) there is seldom any pain resulting from the mismanagement of personal data or outright data disclosure. When a company loses customer data, they face fines, loss of consumer confidence (and hence must spend money to boost advertising), even civil or criminal suits. Both Target Inc (in the USA) [2] and Vodafone (in Greece) [3] after falling victim to cyber attacks that resulted in loss of customer data (and telephone wiretapping in the latter case), suffered serious financial losses. By contrast when the US government surrendered customer data from airline companies such as Jet Blue [4] to a private contractor who posted them on the Internet, or when the UK tax service lost two CDs with 25 million tax payer records, there was little financial or other loss. [5]

To make matters worse, governments, in their fight against tax evasion, are discovering the benefits of gaining access to private data bases such as credit card records, utility billing records and so on. For example, excessive water consumption may be an indication of the presence of a swimming pool which in some countries is taxed as a luxury, or consumption of electricity in a property reported as unoccupied (for tax purposes) may indicate that the owner is renting the property and not declaring the income.

Here we will show how, through the use of heuristics and round-the-clock harvesting of publicly available information, we managed to extract significant portions of databases used by the Greek national health service, the tax authorities and the voter registration database maintained by the Ministry of the Interior.

In all our efforts we only used freely available data, and we did not in any way attempt to "hack" our way. Criminals can obviously take shortcuts and recently a person was arrested with a DVD containing a full copy of the internal database of the Greek tax authority which contains far more information about the Greek taxpayers than anything we could gather though legal means.

Specifically, our work makes the following contributions:

- We have looked at the effect of "joining" disparate publicly available databases on privacy. In addition, we examined how automated exhaustive search attacks against public (i.e. operated by the state) services can uncover private personal data. For example, by trying sequential guesses on the year of birth we can, with negligible effort, since the search space of years of birth is very small, trick one system into revealing the correct guess. By leveraging the ability of networks of computers to mount distributed mass-scale crawling on various data sources we have demonstrated that even entire databases can be downloaded without raising any alarms.
- We demonstrated that uncoordinated web services combined with ad hoc data protection mechanisms can lead both authorities and citizens into a false sense of security. This, in turn, has led to increased efforts to widen the scope of data gathering, causing more serious or extensive leaks. We endeavor to raise the awareness on the part of all relevant stakeholders by demonstrating the scope of such attacks.
- Finally, we demonstrated that natural language documents, especially legal documents, that have a lot of internal structure and standardized means of expression, even scanned ones that need to be OCRed, do not present a significant barrier to automated harvesting of personal information.

Although practically all the personal information we have gathered can be collected from open databases, we have anonymized most of the results presented in this paper (e.g. by smudging national id-card numbers) so as not to cause further damage to the privacy of individuals that are used in examples.

## 2. Public Data Sources

Greek government has offered a variety of data to the public, either to accommodate taxpayers' needs or in terms of transparency. Here we present the key data repositories, their contents and authentication methods..

### 2.1 The Greek Tax Registration Number

Greek Tax Registration Number (in Greek,"ΑΦΜ"), is a unique number provided by the Greek Ministry of Finance to every person or legal entity engaging in financial activity in Greece. Every transaction with public sector services, such as tax offices, require at least this number. The ubiquitous nature of Greek Tax Registration Number (TRN), made other institutions, such as banks or insurance services to incorporate it in their transactions as an additional means of identification.

The Greek TRN consists of 9 digits. It is non-sequential and includes a guard digit which is computed by an algorithm so as to allow the detection of mistyped TRNs and may be found on government web sites.

Greek Tax Registration Numbers are separated in two main categories. The first category includes numbers issued to wage earners and pensioners, while the second category includes self-employed individuals, legal entities, organizations, institutions and businesses. In an effort to aid transactions among entities belonging in the second category, the General Secretariat of Information Systems (GSIS) [6], responsible of computerization of the public sector, introduced a web service that, given an TRN that belongs to the second category, provides an abbreviated version of the record of the entity which matches that TRN. The information provided includes name and commercial title of business, location of headquarters, telephone number as well as the type of business activity [7]. A detailed list of information provided by the service, is presented on Table 1.

| **Record returned by the Tax Registration Number web service** |
|---|
| Tax Registration Number (TRN) |
| Last name, First name and Father's name |
| Commercial title |
| Address (Street, Number, Postal Code, City) |
| Registration Date (Start Date) |
| Stop Date (in case of ex-businesses) |
| Phone & Fax numbers |
| Business Activity |
| Physical or non-Physical Entity Indication |
| Taxation Bureau where business is registered to |
| Active/Inactive TRN Indication |
| Business/Ex-Business Indication |

Table 1: By submitting a valid TRN to the Tax Registration Number web service we receive the above information

If the submitted number does not belong in the second category, the system returns an error which indicates the reason for the refusal, such as unregistered Tax Registration Number, number belongs to a wage earner or pensioner (and hence information about that TRN cannot be released), and so on. This information, as we shall see later, can be used to determine which TRNs to look for in other on-line data sources.

Typically, the TRNs of persons are placed in the first category (i.e. their personal data is not made available), unless they boceme self-employed. In this case, their TRN is moved to the second category where it stays practically for ever.

**2.2 Greek Identification Card Number**

Along with the Tax Registration Number, every Greek citizen is issued with an Identification (ID) card. It holds information about its owner, namely first name, last name, father's name, date of birth, as well as a unique Identification Card Number.

Each Identification Card Number (ID) consists of 1 or 2 Greek capital letters and a 6-digit sequence. The ID card number is the primary form of person identification, and is widely used by whenever a government ID is required. Unlike the social security number and the Tax Registration Number, the Identification Card Number identifies the ID card, not the person, hence changes each time a new ID card is issued. This means that there may well be multiple ID card numbers corresponding to a given person.

In many cases organizations such as banks, request the TRN along with the ID card number, implying the existence of multiple redundant databases in both the private and the public sector that can match TRNs to ID card numbers and vice versa.

**2.3 Greek Social Security Number**

In 2009, the Greek state, introduced the Social Security Registration Number (AMKA) aiming to unify transactions among insurance and healthcare institutions in Greece. Every Greek national is required to have an AMKA number.

AMKA is a structured number consisting of 11 digits. The first six encode the person's date of birth while the following 4 is a sequence number. The last digit usually indicates the sex of the owner. In particular, men are identified using odd digits, while women are identified with even digits [8]. During registration for an AMKA, the Identification Card Number is required. Optionally, the site asks for the TRN [9].

Individuals can find their AMKA, by supplying their first name, last name, father's and mother's name as well as the full date of birth to a government operated website [10]. If the supplied information does not match exactly the corresponding information in the AMKA database, the site returns a message to the effect that the person does not exist. The only exception is the date of birth. If only the year of birth is supplied, the site may ask for additional information (such as person's TRN or Identification Card Number) towards validation. If all information is correct, the AMKA of the person is returned.

The mandatory use of the AMKA in any insurance related claim, provided the Greek Tax authorities, the opportunity to confirm the existence of dependent children, by requiring parents claiming tax deduction because they have underage children to supply the AMKA numbers of their children. Combined with the collection of TRNs by the AMKA site, we observe that there exist at least two datasets that can map from one identifier to the other.

### 2.4 Greek Voter Registration

The Ministry of the Interior, operates a web service to help voters find their electoral center on election day [11]. Voters, enter their first name, last name, father's name, mother's name and their year of birth and get back a screen with information from the Registrar General specifying their assigned electoral center. While operating this service only makes sense during elections, it is available continuously.

### 2.5 Governmental Documents (Diavgeia)

From October 1, 2010, any organization receiving funding from the Greek government, is obliged to upload decisions and other documents to a publicly available repository, called Diavgeia [12] (in Greek, Διαύγεια which means transparency).

According to Diavgeia statistics [13], so far, more than 8.5 million documents have been published by approximately 3,500 public organizations. Documents include hiring or purchasing decisions, detailed payroll lists, balance sheets and so forth. Hence it is an obvious source of private data such as names, Identification Card Number, Tax Registration Number, etc.

## 3. Data Crawling Infrastructure

As we have noted in the previous section, a lot of private data is held in publicly accessible repositories protected with weak authentication mechanisms. The common assumption is that only the owner of this information can access it, because the systems rely on knowledge of personal information that only the owner can know.

This assumption can be broken by, on one hand, searching for the required information in other sites, and by brute force (trying out repeated guesses until we hit the correct value). In some cases, such as the year of birth, the search space is so tiny that the search can be performed manually, in others, such as the exact date of birth, it is far more effective to automate the process. In this section we describe various techniques used and methodology followed towards collecting data from a variety of sources.

Our first target was the Tax Registration Number web service provided by the General Secretariat of Information Systems, described in the previous section. Our objective was to submit all possible TRNs and thus retrieve all the records that the system was willing to supply. In other words we wanted to create a copy of the part of the TRN database that was accessible to the public.

We firstly generated every valid Tax Registration Number. This resulted in approximately 90 million valid Tax Registration Numbers. For each one, we queried the Tax Registration Number web service and recorded the returned information.

To make our collection mechanism more efficient, we segregated the set of valid Tax Registration Numbers into 10 subsets using the first digit as a filter. Each subset contained approximately 9 million elements. As Greece's population does not exceed 11 million residents, it is obvious that the search space would be sparse. However, the allocation of TRNs reveals clustering, hence TRNs beginning with digits 0, 1, 2 and 99 are more likely to correspond to registered TRNs. Thus, we requested numbers from these ranges first. This provided us with high initial yields. Once the populated ranges were mined out, we were less concerned about being locked out of the system and we continued our search into the remaining ranges until we made queries for the entire 90 million of possible TRNs.

To avoid triggering any detection mechanism we implemented the collection mechanism in a distributed fashion and avoided requests on sequential TRNs. A searcher component performs requests to the service using a set of TRNs, while results are stored in a centralized machine. Searchers were deployed on multiple PlanetLab nodes located in Greece.

The Diavgeia document repository was our second target. It employs an indexing mechanism to various document metadata. Unfortunately, it cannot perform searches on document's contents. Luckily, we found another service called yperdiavgeia [14], which indexes all documents that are posted on Diavgeia. This web service incorporates faster search mechanisms as well as applying OCR techniques to scanned documents, broadening the scope of potential results.

In order to automate the process of locating and extracting Identification Card Numbers that appear in documents we used the following heuristic. Many instances of Identification Card Numbers are preceded by the string "AΔT" (the initials of Identification Card Number in Greek). Moreover, we observed that the person's name, surname usually follow the Identification Card Number in the text. We, therefore, searched the yperdiavgeia documents for instances of the string "AΔT" and downloaded them locally. Using pattern matching we identified each Identification Card Number location within the document, and tried to find person's name that may have been located near the ID reference via common syntax patterns. A valid reference to a name along with an Identification Card Number would be "John Papadopoulos, son of George, with AΔT AB-123456".

This heuristic produced very good yields, because most official documents follow set patterns for stating the name, Identification Card Number, and address of individuals.

To crawl the Greek Voter Registration site, we relied on names collected from the previously crawled Tax Registration Number web service. In particular we used 1.9 million records containing first name, last name and father's name as input.

The Greek Voter Registration service requires at least the first two letters of first name, father's name and mother's name. Thus, we brute-forced mother's name using a corpus of most common Greek female names and generating 2-letter combinations. We also brute-forced the year of birth. Since every Greek citizen can vote after being 18 years old, we only needed to search for years in the range between 1912 and 1995.

The data collected from the Greek Voter Registration service were then used to query the AMKA database which uses a more strict authentication mechanism requiring the full date of birth (in contrast to the Greek Voter Registration site which wants only the year of birth). Since we had collected from the Voter Registration site all the required information for each person, including the year of birth, but not the exact date, we needed to guess the correct date of birth from the 366 possible dates.

To optimize the search, we relied on the birthday paradox, thus, reversed the search pattern and starting with January 1st, we checked all the records we had collected from the Greek Voter Registration site. We then removed from the list the records that matched (i.e. the persons that were born on Jan 1st), and continued with January 2nd. Thus our list of persons decreased as we moved later in the year. This approach is more efficient compared to the naive approach of scanning possible dates for each individual. Requests were issued using 500ms interval between them to avoid being potentially blocked from the website.

## Putting it all together

As we described in the introduction, the real threat arises from the combination of these repositories, as the attacker uses one site against the other, so to speak. For example, consider any Greek national. We can locate our victim from any one of a number of sources. We can use the TRN database to locate our victim, or we can use Diavgeia.

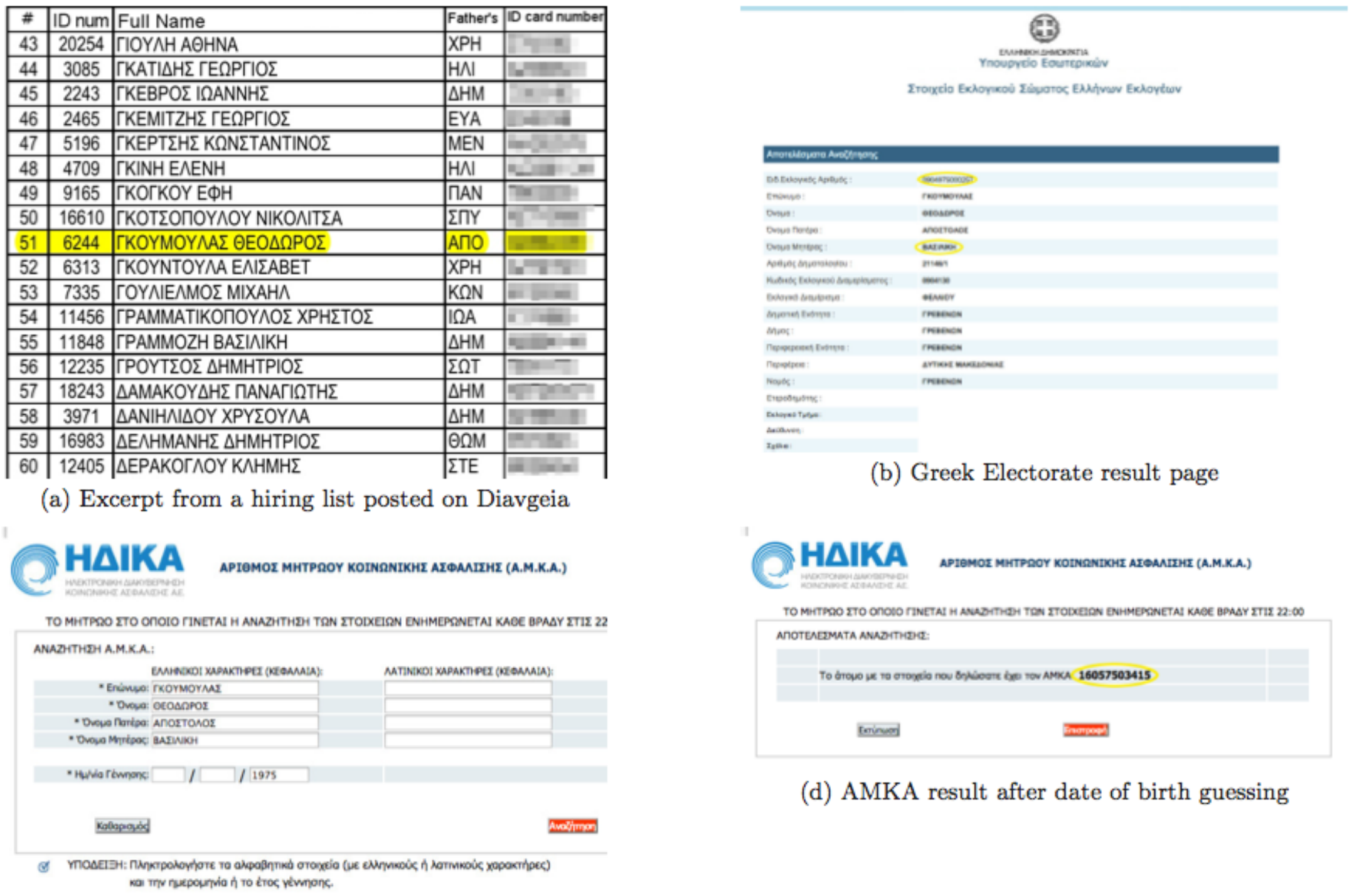

| # | ID num | Full Name | Father's | ID card number |
|---|---|---|---|---|
| 43 | 20254 | ΓΙΟΥΛΗ ΑΘΗΝΑ | ΧΡΗ | [illegible] |
| 44 | 3085 | ΓΚΑΤΙΔΗΣ ΓΕΩΡΓΙΟΣ | ΗΛΙ | [illegible] |
| 45 | 2243 | ΓΚΕΒΡΟΣ ΙΩΑΝΝΗΣ | ΔΗΜ | [illegible] |
| 46 | 2465 | ΓΚΕΜΙΤΖΗΣ ΓΕΩΡΓΙΟΣ | ΕΥΑ | [illegible] |
| 47 | 5196 | ΓΚΕΡΤΣΗΣ ΚΩΝΣΤΑΝΤΙΝΟΣ | ΜΕΝ | [illegible] |
| 48 | 4709 | ΓΚΙΝΗ ΕΛΕΝΗ | ΗΛΙ | [illegible] |
| 49 | 9165 | ΓΚΟΓΚΟΥ ΕΦΗ | ΠΑΝ | [illegible] |
| 50 | 16610 | ΓΚΟΤΣΟΠΟΥΛΟΥ ΝΙΚΟΛΙΤΣΑ | ΣΠΥ | [illegible] |
| 51 | 6244 | ΓΚΟΥΜΟΥΛΑΣ ΘΕΟΔΩΡΟΣ | ΑΠΟ | [illegible] |
| 52 | 6313 | ΓΚΟΥΝΤΟΥΛΑ ΕΛΙΣΑΒΕΤ | ΧΡΗ | [illegible] |
| 53 | 7335 | ΓΟΥΛΙΕΛΜΟΣ ΜΙΧΑΗΛ | ΚΩΝ | [illegible] |
| 54 | 11456 | ΓΡΑΜΜΑΤΙΚΟΠΟΥΛΟΣ ΧΡΗΣΤΟΣ | ΙΩΑ | [illegible] |
| 55 | 11848 | ΓΡΑΜΜΟΖΗ ΒΑΣΙΛΙΚΗ | ΔΗΜ | [illegible] |
| 56 | 12235 | ΓΡΟΥΤΣΟΣ ΔΗΜΗΤΡΙΟΣ | ΣΩΤ | [illegible] |
| 57 | 18243 | ΔΑΜΑΚΟΥΔΗΣ ΠΑΝΑΓΙΩΤΗΣ | ΔΗΜ | [illegible] |
| 58 | 3971 | ΔΑΝΙΗΛΙΔΟΥ ΧΡΥΣΟΥΛΑ | ΔΗΜ | [illegible] |
| 59 | 16983 | ΔΕΛΗΜΑΝΗΣ ΔΗΜΗΤΡΙΟΣ | ΘΩΜ | [illegible] |
| 60 | 12405 | ΔΕΡΑΚΟΓΛΟΥ ΚΛΗΜΗΣ | ΣΤΕ | [illegible] |

(a) Excerpt from a hiring list posted on Diavgeia

(b) Greek Electorate result page

(c) AMKA form

(d) AMKA result after date of birth guessing

**Figure 1:** Attack scenario: The Greek government regularly posts on the Diavgeia web site lists of persons hired by the Greek government. We, can therefore, find from such lists the initial information we require (first name, last name, father's first name) as shown in subfigure (a). We then use other government websites such as the voter registration site (subfigure (b)) and the AMKA site (subfigure (c)) to leverage more information about the person, until we create a complete profile which includes, in addition to our original information, full date of birth, mothers name, voter registration number, AMKA number and so on.

In the example shown in Figure 1, we arbitrarily chose an individual from a hiring list posted online, and managed to reveal his AMKA, mother's name and date of birth, thus creating a "profile" for him.

If this person was self-employed at any time in his life, we can use the TRN database to find his or her tax registration number and home address.

## 4. Is it only a Greek problem?

We used the Greek case as an example of the dangers of personal information leakage from publicly accessible databases, but this is by no means an isolated case.

In many other countries the same problems exist although the means to perform the data mining are different due to differences in organization, authentication, and the techniques and technologies used in the data repositories themselves. We will briefly look at a number of countries that we have looked at in the past couple of years.

### 4.1 United States

Unlike most countries, the US does not have a standardized identification number or document. The most common replacement is the driver's license (for id) and the social security number (SSN). The generalized use of the SSN in conjunction with the fact that the government issued SSN card does not include the photograph of the owner, means that using someone else's SSN number is relatively easy. Although for native born US residents the SSN is roughly correlated with their age, this is not the case for persons who arrive in the US later in life.

Reports published each year by the Identity Theft Resource Center show a growing number of instances where another person's SSN is used in the filing of tax returns (in most cases so that the fraudster can collect a tax reimbursement), in the issuing of a driver's license, and in applications for emplsoyment (either by persons who would normally be disqualified for the particular job because of e.g. their background, or by illegal immigrants who cannot get a SSN).

The structure of the Social Security Number is nine digits that can be divided intro three distinct parts. The first three numbers is called area number, and can potentially uncover the previous/current residence of the person. In general, numbers were assigned (in 25 June 2011, SSA changed the SSN assignment process to "SSN randomization") beginning in the northeast and moving south and westward, so that people on the East Coast had the lowest numbers and those on the West Coast had the highest numbers. The middle two digits were the Group Number and the last four digits were Serial Numbers that represent a straight numerical sequence of digits from 0001 to 9999 within the group. A 2009 paper by Allessandro Acquisti and Ralph Gross describe how SSNs can be predicted from publicly available data such as an individual's place and date of birth. Leveraging the structure of the SSN, the authors combine information from multiple sources including the Social Security Administrations Death Master File, data brokers, and profiles on social networking sites to infer the SSN of randomly selected individuals. This highlights the additional danger that in impostor can select an SSN that is closer to his or hers background and hence less likely to attract additional scrutiny.

Although the private data provided by the electronic services seems to be secure enough, there are many cases that SSNs leak from reports issued by federal agencies. Agencies generally place no restrictions on the reuse of data included in public records. This means that information can change hands many times and even be outsourced to foreign service providers. A U.S. Government Accountability Office's (GAO) report[1] spublished in 2006 and entitled "Social Security Numbers Are Widely Available in Bulk and Online Records" reports that SSNs are displayed on millions of cards issued by federal agencies, including 42 million Medicare cards, 8 million Department of Defense identification cards and insurance cards, and 7 million Veteran Affairs identification cards. However they claim that some States try to mitigate this leak by enacting laws to restrict the use and display of SSNs. Such belated efforts, however, do nothing to retrieve the millions of SSNs already available through public records

[1] http://www.gao.gov/new.items/d081009r.pdf

### 4.2 Brazil

In Brazil the official ID card is associated to a number called *Registro Geral* (RG). Although the ID cards are supposedly national, the RG numbers are assigned by the states and a few other organizations, such as the armed forces. RG numbers can have verification digits or letters and each state can design its own system. So, not only is it possible for a person to have the same RG number as a person from other state, but it is also possible to have more than one RG, from different states. Each RG card also includes another number, called *Cadastro de Pessoas Físicas* (CPF) that is also used as a proof of identification. CPF is a federal and unique eleven-digit number that it was created originally for purposes of taxation. Apart from RG and CPF there is also a third identification number that is used for companies, named *Cadastro Nacional de Pessoas Jurídicas* (CNPJ).

In order to find out if Brazilian federal government includes private data in public reports, we used as a test case a well known fiscal transparency portal named Portal da Transparência[2]. This portal was started in 2004 and draws together information from across the federal government. The portal includes a large amount of information on public revenues, budgets and spending as well as income and assets of public servants. It is updated daily and provides searchable, open access to the data. During our examination we focused on financial reports from the latest Football World Cup. From these reports we saw that although the reports contained many details about employees (such as Name, Job, Working Hours, Job Location, Date of Hiring, etc.) the CPF number was always truncated (5 last digits were missing). This means that the government is able to successfully protect their citizens' data when it publishes data for fiscal transparency purposes.

We then tried to see if a user is able to gain access to private data of other users by using the electronic services of the Ministry of Finance and the Department of Federal Revenue of Brazil.[3] From this portal a user by providing a CPF, a Date of Birth and filling a captcha can generate an Access Code. With this Access Code, the CPF and a password or just his Digital Certificate he can login to his account in e-CAC. e-CAC (Virtual Taxpayer Service Center) is an electronic portal where various services protected by tax secrecy can be performed via the Internet by taxpayers themselves. These services allow a user to: check for any disputes in the Declaration of Personal Income Tax, obtain copies of statements, rectify payments, check partial debts, print the proof of registration with the CPF, access to his inbox where he can access notices sent by the IRS. However, besides the strict access control that e-CAC adopts in order to secure its important services and the sensitive data these services handle, there is a small web service[4] in the Ministry of Finance's portal mentioned above that takes as input a CPF and a captcha and gives as an output a Registration Status Proof of CPF. This Registration Status Proof, among other information, includes also the Full Name of the CPF holder and the type of the Registration. To make matters worse, we found an online CPF number generator and hence, we were able to successfully retrieve several CPF and Full Name pairs.

---

[2] www.portaltransparencia.gov.br

[3] http://idg.receita.fazenda.gov.br

[4] http://www.receita.fazenda.gov.br/Aplicacoes/ATCTA/CPF/ConsultaPublica.asp

### 4.3 Malaysia

Malaysia is the first country in the world to introduce an identification card that incorporates both photo identification and fingerprint biometric data on an in-built computer chip embedded in a plastic card. This plastic card, introduced in 2001, is named MyKad and it is the compulsory identity document for any citizen aged 12 and above. MyKad is designed to provide some significant functions such identification, driver's license, travel document, health information (blood type, allergies, chronic diseases, etc.), e-cash function, ATM integration, transit card (Touch N'Go), and digital certification known as the Public Key Infrastructure for e-commerce transactions. Users can access personal information on their cards at government kiosks and offices, after biometric authentication of their fingerprint and providing MyKad'ss PIN. Access to personal information by others is hierarchical, for example, only certain medical officers have access to sensitive health information. However, access to some personal information held in the MyKad system seems to be available, remotely via a network, to a wide range of third parties, including hotels, restaurants and ticket agents. In general, the wealth of personal information available on the MyKad has raised concerns pertaining the privacy risks and possible exploitation of the easily accessed data. The Federation of Malaysian Consumers Associations has also expressed concern over the government's lack of implementation of clear guidelines or consultation with the public on how the MyKad is to be used, by whom, and for what purpose. In general, Malaysians cannot be certain who has access to what information and for what purpose. On MyKad there is a National Registration Identification Card Number (NRIC No.). NRIC is introduced in 1990 and is a unique 12-digit number issued to Malaysian citizens and permanent residents. According to MyID, a Malaysian Government initiative, the NRIC Number is the sole reference number for Malaysians in their transactions as an individual with the government agencies. NRIC number uses a YYMMDD-SS-###G format. The first group of numbers (YYMMDD) are the date of birth. The second group of numbers (SS) represents the place of birth of the holder. The last group of numbers (###G) is a randomly generated number. The last digit (G) is an odd number for a male, while an even number is given for a female. Thus the structure of the NRIC allows a third party to deduce personal information the gender of the holder and the date and the place of birth.

By searching the web for potential NRIC number leaked from public articles and reports, we found large numbers of governmental reports that included unprotected NRIC numbers. One example is the report of the Malaysian Electoral Roll Analysis Project (MERAP) in 2012 that analyzed the electoral process, while another is the government's web site[5] that publishes not only the general results of the parliamentary elections and the State assembly from several years, but also the NRIC number of every candidate. Collecting all these leaked NRIC numbers and with the help of a web application[6] hosted in the same web site, by giving the NRIC number as an input anyone can retrieve the holder's Name, Birth Date, Gender, Voting District, and State.

[5] http://semak.spr.gov.my/spr/laporan/5_KedudukanAkhir.php

[6] http://daftarj.spr.gov.my/DAFTARJ/DaftarjBM.aspx

### 4.4 India

In India, although various schemes for a national identification number have been proposed, the closest India has come to this is the Permanent account number (PAN), that is a unique, 10-character alpha-numeric identifier, in the form of AAAPL1234C, issued by the Indian Income Tax Department under the supervision of the Central Board for Direct Taxes (CBDT), for purposes of tracking income and income taxes.

Every receiver of income with taxable income has to have a PAN and in recent years has to be produced practically for every financial transaction. While the alphanumeric PAN number is unique, individuals and corporate entities have been able to obtain multiple PAN cards fraudulently. According to official data, there are only 30 million income tax payers in India, yet there are 170 million genuine PANs issued as of 2014.

Under the same Indian Income Tax Act, there is also a Tax Deduction and Collection Account Number (TAN). TAN is a 10 digit alphanumeric number issued to persons who are required to deduct or collect tax on payments made by them.

Every Indian citizen is able to have electronic access to information and services at the official portal of Income Tax Department of the Ministry of Finance of the Government of India.[7] These services include electronic filing of income tax return and forms, services like "Know Your PAN", "Know Your TAN", "Know Your Jurisdiction", viewing tax credit statement, filing of income tax return through e-Return intermediary and many more. By examining the existing level of protection of the data stored in this portal we saw some interesting facts. More specifically, in "Know Your PAN" service a user, by providing a date of birth, a surname and filling a weak numeric captcha, is able to retrieve everyone's PAN. Moreover, in "Know Your TAN" service a user can provide a Category of Deductor (Branch of Company, Company, Individual, Government etc), a State and either a TAN or a Name and, after filling the weak numeric captcha, he can get all the TAN records in a specific state. For example we choose the Individual category, the Delhi State, we fill in the name "John" and we get all the TAN records of all the people in Delhi State named "John". To make matters worse, the information published inside these TAN records are: Full Name, Address, ZIP code, email address, TAN and PAN. Considering the many uses of PAN in financial transactions and identification processes in India we can easily identify the importance of this major sensitive data leakage from the Indian government itself.

### 4.5 Singapore

In Singapore the identity document in use is The National Registration Identity Card (NRIC) and it is compulsory for all Singaporean citizens and permanent residents who are fifteen years of age and older. On the front side of the card, are the holder's name, race, date of birth, sex, country of birth, and a colour photograph. On the back of the card is the NRIC number and its bar code, a fingerprint, issue date of the card, and the holder's current residential address. In Singapore simply providing an NRIC number without producing the card itself will suffice for verifying a person's identity. The NRIC number consists of total 9 characters in the pattern #0000000@, where "#" is a letter that can be “S”, “F”, “T” or “G” depending on the status of the holder. Citizens' and permanent residents' identity number starts with prefix S if they are born before 2000 and T if they are born in or after year 2000. This character is followed by a 7-digit number and a checksum alphabet. For citizens and permanent residents born after 1968, the first two digits of the 7-digit number indicate their birth year. There are also several web applications that try to figure out the holder's month of birth from a given NRIC number.

---

[7] https://incometaxindiaefiling.gov.in/e-Filing

From July 2014 the Personal Data Protection Act (PDPA)[8] came into force requiring organizations in Singapore to comply with specific data protection provisions including who has the right to use and publish NRIC numbers and other private data, and how these sensitive data must be protected in case of publication.

However, there is no action for already published NRIC numbers. This way we succeed in retrieving some NRICS from documents published by the High Court of the Republic of Singapore.

In order to check for sensitive data leaks from the Singaporean governments' web sites we visited the Inland Revenue Authority of Singapore (IRAS).[9] This is a web site for taxation services, that includes myTax Portal,[10] a portal that allows all taxpayers to transact with IRAS electronically. We were unable to access any e-service from IRAS site because in order for a user to use a e-service he has to be authenticated first by using a NRIC number along with an IRAS PIN number or a password named SingPass that is a common password for all government web services.

## 5. Mitigation

The problem of balancing transparency against the protection of privacy is very hard indeed, and to a large extent philosophical, rather than technical. Nevertheless, there are numerous techniques that, if they have been implemented, would have radically reduced the extent of the leaks by limiting the effectiveness of the brute force attacks we have carried out as part of our research. In this section we discuss countermeasures that could be adopted by government entities towards prevention of similar information leaks in the future.

- **Rate-limiting** Rate-limiting techniques are widely used to throttle the number of requests originating from a specific user or host. Despite its primary use in thwarting Denial-of-Service (DoS) attacks, rate-limiting can be also used to prevent rapid-fire requests of the type that we described earlier in this paper. For example, the TRN web service can introduce a daily limit on the number of requests that can be issued from a given IP address. The granularity of this limit can be adjusted to accommodate legitimate uses of the facility. For instance, the IP addresses of accounting offices or other organization that may need to perform very large numbers of requests daily may be granted a higher daily limit.

---

[8] http://www.pdpc.gov.sg/docs/default-source/publicconsultation/ advisory_guidelines_on_selected_topics.pdf?sfvrsn=2

[9] http://www.iras.gov.sg

[10] https://mytax.iras.gov.sg

From our crawling experience, we faced lock-outs from the TRN web service after a long period of crawling. Lock-outs were based on the IP address used. We are not aware as to whether this lock was manually enforced, or automatically, triggered by a rate limiting mechanism.

Since we were not the only ones downloading the contents of that particular service (we are aware of at least one company which apparently did something similar) we were not surprised when the service was eventually discontinued temporarily, with no specific justification.

- **CAPTCHA** Along with rate-limiting techniques, websites having forms, such the Greek Voter Registration site [15], can also be enforced with CAPTCHA methods. After a number of successive requests originated from the same IP address, a CAPTCHA must be solved. This would spurn most automated brute forcing attempts. During crawling, none of the websites had incorporated CAPTCHA techniques.

  For some web services, such as the TRN service, the root of the problem was lack of any user authentication. By authenticating users and asking them to agree to some guidelines on the use of the service, there would be little scope for the kind of massive data downloads we have carried out. In many cases, the simple fact that the client has been identified is sufficient to deter abuse. For the TRN service in particular, the only justification for not limiting the use of the service to Taxis users, is probably the requirement that users from other EU countries be able to check TRN numbers.

- **Data Sanitization** As we have observed, much sensitive information was hidden inside Government documents found either on the Diavgeia repository or on municipality websites. With the introduction of the yperdiavgeia search engine [16], this unstructured information can be indexed. Thus an attacker may search for specific names or Tax Registration Numbers of his interest, performing a more targeted attack.

  As a countermeasure, Diavgeia document repository can sanitize references to names or surnames prior to document publication. Instead of displaying the full name of an individual, only a portion may be visible. For example, a reference to "John Papadopoulos" would be sanitized to "J. Papad.".

  Moreover, governments can enforce a stricter policy, for making Personally Identifiable Information (PII) available from sources outside Diavgeia. As we have shown, major privacy leaks were effected though municipal or other institutions linked to the public sector. Decisions containing sensitive information must be sanitized or anonymized and sent to Diavgeia.

- **Accountability** By identifying civil servants who are responsible of PII leaks, we envisage that proper vigilance will be observed on the part of the authorities who publish documents on public websites. To that effect, analysis of document metadata (e.g. Word documents storing the name of the author or the modification date) may produce valuable information leading to the source of disclosures of PII to the public [17,18].

Finally, Government can also use decoy documents [19], with “bait” information like TRNs or AMKAs, as a method to identify leaks.

## 5. Conclusion

In this paper we discussed ways that information from multiple on-line government data repositories may be combined to create profiles of Greek citizens.

As this sensitive information is publicly available, a miscreant may exploit it for malevolent purposes. Furthermore, both legal and ethical questions arise from the large scale availability of this type of sensitive information.

There is, clearly, a need for organization and privacy awareness. However, the problem  cannot be solved by technical means, the mentality of the civil servants needs to change. Unfortunately, civil servants, owe allegiance to their service and not to the citizens - customers. Hence they are more likely to ignore privacy guidelines, believing, rightly or wrongly, that their service will protect them from any legal consequences that their actions may have. It is, therefore, imperative to resist the expansion of government administered data repositories until sufficient safeguards for their correct and privacy preserving operations can be established.

## References

1 www.wired.com/2012/08/apple-amazon-mat-honan-hacking
2 http://www.huffingtonpost.com/tag/target-hacked/
3 Vassilis Prevelakis and Diomidis Spinellis, “The Athens Affair,” IEEE Spectrum, **44**(7), pp. 26-33, July 2007.
4 http://archive.wired.com/politics/security/news/2003/10/60885?currentPage=all
5 “UK Government disks were not well encrypted,” comp.risks **24**(92).
6 General Secreteriat for Information Systems. www.gsis.gr
7 VAT Registration Numbers Web Service. http://www.gsis.gr/wsnp/wsnp.html.
8 on AMKA - blog.postmaster.gr. http://blog.postmaster.gr/2009/05/20
9 http://www.amka.gr/odigos4.html
10 AMKA Web Service. https://www.amka.gr/AMKAGR
11 Greek Electorate Web Service. http://www.ypes.gr/services/eea/eea.htm
12 Diavgeia Document Repository. http://diavgeia.gov.gr
13 **«SAME AS 12»**
14 UltraCl@rity - Search in the depths of the Cl@rityprogram. http://www.yperdiavgeia.gr.
15 **«SAME AS 13»**
16 **«SAME AS 20»**
17 T. Aura, T. A. Kuhn, and M. Roe. Scanning Electronic Documents for Personally Identifiable Information. In Proceedings of the 5th Annual ACM Workshop on Privacy in the Electronic Society, pages 41–50. ACM, 2006
18 E. Gessiou, S. Volanis, E. Athanasopoulos, E. P. Markatos, and S. Ioannidis. Digging up Social Structures from Documents on the Web. In Proceedings of the Global Communications Conference (GLOBECOM), pages 744–750. IEEE, 2012
19 B. M. Bowen, S. Hershkop, A. D. Keromytis, and S. J. Stolfo. Baiting Inside Attackers Using Decoy Documents. In Proceedings of the 5th International ICST Conference on Security and Privacy in Communication Networks, pages 51–70, 2009